\pdfoutput=1

\documentclass[11pt]{article}

\usepackage[final]{acl}

\usepackage{times}
\usepackage{latexsym}

\usepackage[T1]{fontenc}

\usepackage[utf8]{inputenc}

\usepackage{microtype}

\usepackage{inconsolata}

\usepackage{graphicx}

\usepackage{amsmath}
\usepackage{amssymb}
\usepackage{amsfonts}
\usepackage{mathtools}
\usepackage{dsfont}
\usepackage{relsize}
\usepackage{anyfontsize}

\usepackage{algorithm2e}
\SetKw{Input}{Input}
\SetKw{Output}{Output}
\RestyleAlgo{ruled}

\usepackage{subfigure}

\usepackage{multirow}
\usepackage{booktabs}
\usepackage{siunitx}

\usepackage{pifont}
\newcommand{\cmark}{\text{\ding{51}}} 
\newcommand{\xmark}{\text{\ding{55}}} 

\usepackage{enumitem}

\usepackage{xspace}

\newcommand{\ourmethod}{MixLoRA-DSI\xspace}

\DeclareMathOperator*{\argmax}{arg\,max}

\DeclarePairedDelimiterX{\infdivx}[2]{(}{)}{#1\;\delimsize\|\;#2}
\newcommand{\kldiv}{D_{KL}\infdivx}

\title{MixLoRA-DSI: Dynamically Expandable Mixture-of-LoRA Experts for Rehearsal-Free Generative Retrieval over Dynamic Corpora}

\author{
    Tuan-Luc Huynh\textsuperscript{1} \enskip {\bf Thuy-Trang Vu\textsuperscript{1}} \enskip {\bf Weiqing Wang\textsuperscript{1}} \enskip \\
    {\bf Trung Le\textsuperscript{1}} \enskip {\bf Dragan Ga\v{s}evi\'c\textsuperscript{2}} \enskip {\bf Yuan-Fang Li\textsuperscript{1}} \enskip {\bf Thanh-Toan Do\textsuperscript{1}} \\
    \textsuperscript{1} Department of Data Science \& AI, Monash University, Australia \\
    \textsuperscript{2} Department of Human Centred Computing, Monash University, Australia \\
    \texttt{\{tuan.huynh1,trang.vu1,teresa.wang},\\ \texttt{trunglm,dragan.gasevic,yuanfang.li,toan.do\}@monash.edu}
}

\begin{document}
\maketitle

\begin{abstract}
Continually updating model-based indexes in generative retrieval with new documents remains challenging, as full retraining is computationally expensive and impractical under resource constraints. We propose MixLoRA-DSI, a novel framework that combines an \emph{expandable} mixture of Low-Rank Adaptation experts with a layer-wise out-of-distribution (OOD)-driven expansion strategy. Instead of allocating new experts for each new corpus, our proposed expansion strategy enables sublinear parameter growth by selectively introducing new experts only when significant number of OOD documents are detected. Experiments on NQ320k and MS MARCO Passage demonstrate that MixLoRA-DSI outperforms full-model update baselines, with minimal parameter overhead and substantially lower training costs.\footnote{Our code is available at \href{https://github.com/LouisDo2108/MixLoRA-DSI}{\texttt{https://github.com/LouisDo2108/MixLoRA-DSI}}.}
\end{abstract}
\section{Introduction}
\label{sec:introduction}
Generative retrieval (GR) leverages pretrained Transformer models as model-based information retrieval (IR) index, also known as differential search index (DSI), to encode corpus information directly within the model parameters~\cite{tay_dsi_neurips22}. 
However, GR methods primarily assume a static corpus, where the document set remains unchanged. This overlooks the real-world challenge of IR systems to handle dynamic corpora, where they need to continually integrate new documents.

Since GR methods perform indexing via model training, continually updating index may require expensive retraining.
Recent works have explored continual learning (CL) approaches to mitigate catastrophic forgetting while efficiently updating the model without retraining from scratch~\cite{mccloskey_cl_1989,mehta_dsipp_emnlp23}, such as rehearsal-based CL~\citep{mehta_dsipp_emnlp23,kishore_incdsi_icml23,chen_clever_cikm23}. However, these methods either adopt atomic document identifiers (docids)~\cite{mehta_dsipp_emnlp23} or only evaluate on small-scale datasets, limiting their scalability on large-scale retrieval benchmarks~\cite{bajaj_msmarco_arxiv16}.
Additionally, rehearsal-based methods require access to previous documents, raising privacy concerns~\citep{shokri_privacy_sigsac15}, making them impractical for real-world applications. 
Furthermore, dynamic corpora naturally align with task-agnostic (i.e., task-free) CL, characterized by blurry task boundaries~\cite{aljundi_taskfreecl_cvpr19}, since new documents may share high semantic similarity to previous indexed ones. These challenges collectively indicate the need for a \emph{\textbf{task-agnostic}}, \emph{\textbf{rehearsal-free}}, and \emph{\textbf{scalable}} solution.

\begin{figure}[t]
      \centering
      \includegraphics[width=0.95\linewidth]{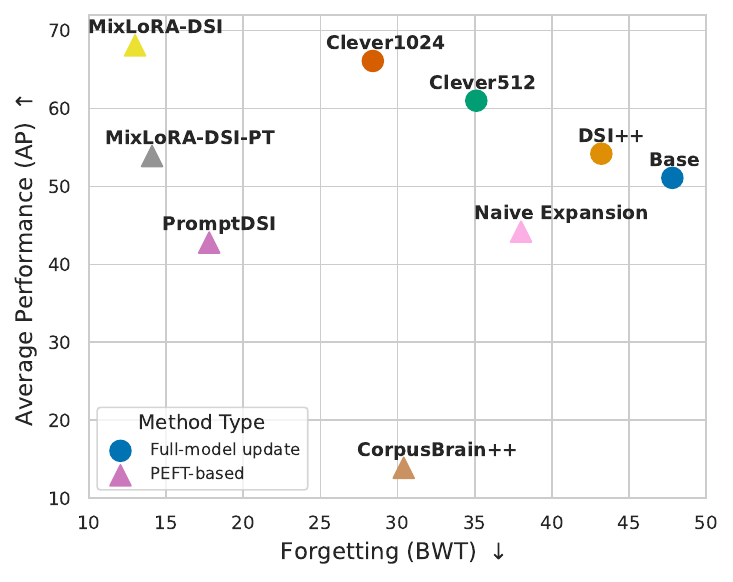}
      \caption{Continual learning performance of different Generative Retrieval methods on NQ320k (Recall@10). PEFT-based MixLoRA-DSI variants offer the best trade-off between average performance and forgetting, while being significantly more parameter-efficient than full-model updates. See Table~\ref{tab:nq320k_main_result} for more detailed results.}
      \label{fig:1}
\end{figure}

In this work, we propose \emph{\textbf{\ourmethod}}, a parameter-efficient fine-tuning (PEFT)-based {dynamically expandable framework for rehearsal-free GR over dynamic corpora}, which outperforms various GR methods (Figure~\ref{fig:1}). 
Our approach builds on mixture-of-experts (MoE)~\cite{shazeer_sparsemoe_iclr17}, where each expert is a low-rank adapter~\citep{hu_lora_iclr22}. MixLoRA-DSI replaces the standard top-$k$ linear router with a top-$k$ cosine classifier~\cite{gidaris_cosineclassifier_cvpr18} and introduces a novel auxiliary loss. 
This improved router alleviates the recency bias in MoE routing of the original router, while the novel auxiliary loss encourages accurate token-to-expert alignment and preserves expert diversity.
As a result, \ourmethod is inherently \emph{\textbf{task-agnostic}} as the top-k routers dynamically select experts per token without relying on rigid task-specific parameters.
The use of LoRA experts ensures efficient indexing while preserving knowledge from previously indexed documents, mitigating catastrophic forgetting and enabling \emph{\textbf{rehearsal-free}} CL.
We enhance \emph{\textbf{scalability}} by opting for residual quantization (RQ)-based docids \citep{zeng_ripor_www24} and developing \emph{\textbf{CL strategies tailored to RQ-based docids}} by leveraging their structural properties.

\ourmethod views GR over dynamic corpora through the lens of out-of-distribution (OOD) detection. Naively adding new experts for each corpus update leads to linear parameter growth without ensuring efficient capacity utilization. However, in practice, queries often share structural patterns and differ mainly in \emph{\textbf{OOD keywords}} associated with newly indexed documents, thus new docids may frequently overlap in latent space. Building on this insight, we propose an \emph{\textbf{layer-wise OOD-driven expansion strategy}} that selectively adds new experts when existing ones fail to capture significant novel information from new documents. This decision is governed by the \emph{\textbf{layer-wise energy scores}}~\cite{liu_energyood_neurips20} from the routers, enabling sublinear parameter growth while maintaining retrieval effectiveness. Overall, we make the following contributions:

\begin{itemize}[topsep=2pt,parsep=2pt,partopsep=2pt,leftmargin=8pt]
    \item We propose \ourmethod, the first OOD-driven dynamic expansion framework for rehearsal-free GR over dynamic corpora, achieving sublinear parameter growth.
    \item We improve MoE routing by replacing the standard router with a top-$k$ cosine classifier, jointly optimized with a novel auxiliary loss to encourage expert specialization while ensuring balanced token-to-expert assignments. 
    \item We propose continual learning strategies tailored to RQ-based document identifiers.
    \item Experiments on NQ320k~\cite{kwiatkowski_nq320k_tacl19} and MS MARCO Passage dataset demonstrate that \ourmethod is significantly more parameter-efficient and robust to forgetting than full-model update baselines.
\end{itemize}
\section{Preliminary}
\label{sec:preliminary}

\subsection{Generative Retrieval (GR)}
\label{subsec:gr}

\paragraph{Differentiable Search Index (DSI).} 
\label{par:dsi}
DSI~\cite{tay_dsi_neurips22} is a representative method in generative retrieval (GR); we use DSI and GR interchangeably. 
Let \(f_\theta\) denote the T5~\cite{raffel_t5_jmlr20}-based DSI model. 
During indexing, DSI is trained to map each document \(d_i \in D\) to a unique identifier \(id_i \in I_D\). 
During the retrieval stage, given a set of query-docid pairs \(Q^{\text{retrieval}}_D=\{(q_i, id_{q_i})\}\), DSI autoregressively generates a ranked list of top-\(k\) docids using constrained beam search, which are then mapped back to documents in \(D\).

Document representations and document identifiers (docids) are two crucial design choices in DSI. At retrieval time, since DSI must process queries, which are typically much shorter and linguistically distinct from the training documents, it results in a distribution mismatch between indexing and retrieval. To address this issue, the current standard practice is to represent documents using pseudo-queries generated by off-the-shelf query generation models such as docT5query~\cite{nogueira_doct5query_2019}. These pseudo-queries are synthetic queries that simulate what users might ask to retrieve the corresponding documents. They serve as replacements for the original document content during indexing, helping to better align the training and inference distributions~\cite{zhuang_dsiqg_22,pradeep_genirscaling_emnlp23}.
We adopt Residual Quantization (RQ)-based docids, encoding documents as sequences of RQ codes~\cite{babenko_rq_cvpr14}, which is the first to show effectiveness on large-scale IR~\cite{zeng_ripor_www24}.

\paragraph{Dynamic corpora in GR.}
\label{subsec:cl-dsi}
Dynamic corpora~\cite{mehta_dsipp_emnlp23} assumes there are \(T+1\) corpora \(\{D_0, \ldots, D_T\}\), each associated with a set of docids \(\{I_0, \ldots, I_T\}\). 
\(D_0\) is a large-scale corpus with annotated query-docid pairs, while \(D_{>0}=\{D_1, \ldots, D_T\}\) contains newly added documents without annotations.
Let \(\theta_{t-1}\) denote the GR model parameters after indexing \(\{D_0, \ldots, D_{t-1}\}\). 
At each timestep \(t>0\), the model must update to \(\theta_t\) to incorporate \(D_t\). 
Evaluation at timestep \(t\) is performed on retrieval queries \(\{Q^{\text{retrieval}}_{D_0}, \ldots, Q^{\text{retrieval}}_{D_t}\}\) using \(\theta_t\).

\subsection{Mixture of LoRA Experts}
\label{subsec:moe}

\paragraph{Mixture of Experts (MoE).} An MoE layer~\cite{shazeer_sparsemoe_iclr17} comprises of \(N\) feed-forward network (FFN) experts, denoted as \(\{E_i\}_{i=1}^{N}\), and a router \(R \in \mathbb{R}^{\text{dim}\times N}\). Given token \(x\), the router computes logits \(R(x)=R^{\top}x\). The gate value of the \(i\)-th expert is computed as:
\begin{equation}
    \label{eq:router}
    p_i(x)=\frac{e^{R(x)_i}}{\sum_{j=1}^{N} e^{R(x)_j}}, \nonumber
\end{equation} 
where \(p_i(x)\) represents the softmax-normalized probability for expert \(E_i\). The token is routed to a subset of experts corresponding to the top-\(k\) gate logits. Let \(Topk(\cdot)\) be a function that selects the top-\(k\) highest logits while setting the rest as zero. The output of a MoE layer is a weighted sum of the top-\(k\) experts, with the weights determined by the gate values:
\begin{equation}
    \text{MoE}(x)=\sum_{i=1}^{N} Topk(p_i(x)) E_i(x). \nonumber
    \label{eq:moe}
\end{equation}
MoE is often trained with an auxiliary load balancing loss to mitigate the unbalanced load for experts when training~\cite{fedus_switchtransformer_jmlr22}. 
Given $N$ experts and a batch $B$ with $T$ tokens, the auxiliary loss \(\mathcal{L}_{\text{aux}}\) is computed as following:
\begin{equation}
\begin{gathered}
    \label{eq:load_balance_loss}
    \mathcal{L}_{\text{aux}}^\text{original}=a\cdot N\cdot \sum_{i=1}^N \mathcal{F}_i\cdot \mathcal{P}_i, \\
    \mathcal{P}_i=\frac{1}{T} \sum_{x\in B} R(x)_i, \\
    \mathcal{F}_i=\frac{1}{T} \sum_{x\in B} \ \mathds{1}\{\argmax_k R(x)_k=i\},
\end{gathered}
\end{equation}
where \(\mathcal{F}_i\) is the fraction of tokens dispatched to expert \(i\), \(\mathcal{P}_i\) is the fraction of the router probability allocated for expert \(i\), and \(\alpha\) stands for the weight coefficient of this auxiliary loss.

\paragraph{MixLoRA: Mixture of LoRA Experts.}
MixLoRA~\cite{yu_moeloraclvision_cvpr24} employs LoRAs~\cite{hu_lora_iclr22} as experts for parameter-efficient fine-tuning. Specifically, the pre-trained FFN remain frozen, and we apply a unique set of LoRAs to each FFN layer. For all experiments, we employ MixLoRA with top-2 router in specific T5's decoder blocks, since the decoder is responsible for autoregressive docid generation. Our MixLoRA implementation is detailed in Appendix~\ref{appendix:mixlora}.

\section{Method}
\label{sec:methods}

\subsection{\ourmethod}
\label{subsec:mixlora_dsi}

\begin{figure*}[t]
    \centering
    \includegraphics[width=0.95\linewidth]{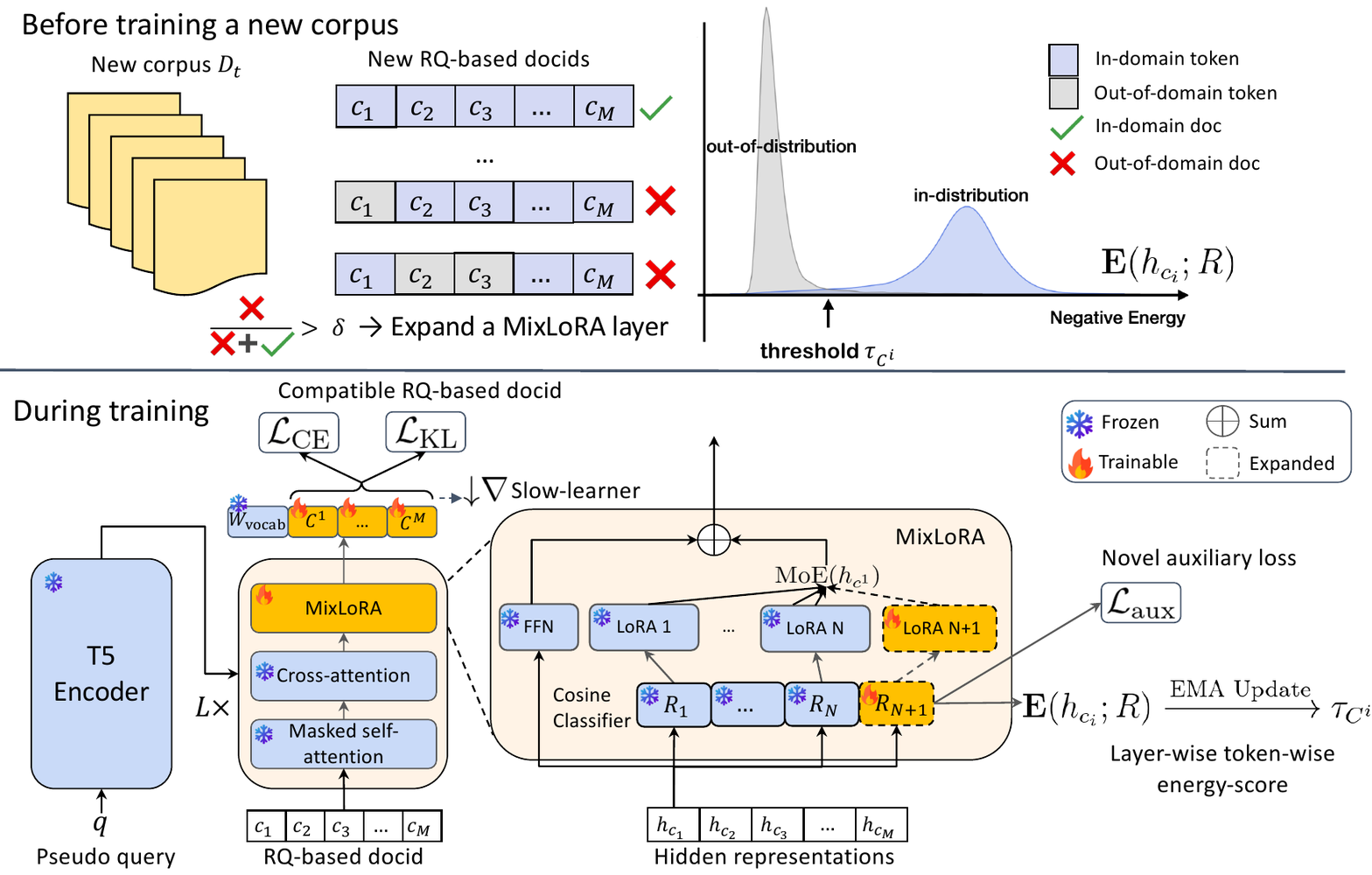}
    \caption{Overview of MixLoRA-DSI. 
    Before training, the model scans the new corpus to to perform energy-based OOD dynamic expansion (Section~\ref{par:energy_ood}). 
    During training, the new router weights are optimized with the novel auxiliary loss \( \mathcal{L}_\text{aux} \) (Eq.~\ref{eq:router_aux_loss}), and the energy score thresholds are updated via exponential moving average (EMA). New LoRA experts and RQ-based docid embeddings \( \{C^1, \ldots, C^M\} \) are trained with the proposed RQ-based docids CL strategies (Section~\ref{subsec:rq_cl_strategies}).
    Residual connections and layer normalizations are dropped without loss of generality.
    }
    \label{fig:mixloradsi}
\end{figure*}

Figure~\ref{fig:mixloradsi} presents the overview of \ourmethod,
a novel dynamic expansion framework for rehearsal-free GR over dynamic corpora, equipped with an OOD-driven dynamic expansion mechanism, IR-specific improved routers, and RQ-based docids CL strategies.

\paragraph{Expanding MixLoRA.}
MixLoRA can be expanded with new LoRAs while maintaining nearly constant computational cost.
Each expansion adds a new LoRA to the existing set and appends a weight vector \(w \in \mathbb{R}^{\text{dim}}\) to the router: \(R^\prime=\{R; w\}\).
Naively adding one expert per corpus to each MixLoRA layer causes linear parameter growth without guaranteed efficiency.
To address this, we seek a principled criterion to determine: \textbf{\textit{When should we expand a specific MixLoRA layer?}}

\paragraph{Energy-based OOD dynamic expansion.}
\label{par:energy_ood}
We observe that:
(1) LoRA experts store knowledge, while the router selects which expert to activate, naturally making it responsible for deciding when expansion is needed. 
(2) In IR, queries often share structural similarities, with differences primarily in OOD keywords. Despite having unique docids, new documents frequently overlap in latent space.
We thus propose a dynamic expansion strategy that detects OOD signals in the latent space relative to previously indexed documents, ensuring expert expansion only when necessary.

Among various OOD detection methods, we adopt the energy score~\cite{liu_energyood_neurips20}, because of its two key advantages:
(1) The router in each MixLoRA layer functions as a classifier, making the energy score a natural fit.  
(2) Unlike generative methods, it avoids computationally expensive and often unreliable density estimation~\cite{lee_cndpm_iclr20,ardywibowo_varigrow_icml22,nalisnick_deepgenood_iclr19}.
Given an input token \(x\) to the router \(R\) of a MixLoRA layer, we treat the router as an \(N\)-class classifier. 
The energy score parameterized by the router \(R\) is computed as:
\begin{equation}
    \label{eq:energy_ood_score}
    \mathbf{E}(x; R) = -T \cdot \log \sum\limits_{i=1}^{N} \exp(\langle R_i, x\rangle / T),
\end{equation}
where \(T\) is the temperature, typically set to 1.0 to ensure the energy score remains parameter-free, and \(R_i\) represents the weight vector of the router corresponding to the \(i\)-th expert.

The energy-based OOD dynamic expansion process is illustrated in the top-half of Figure~\ref{fig:mixloradsi}. We compute energy scores for each token at every MixLoRA layer. During indexing, we maintain exponential moving averages of in-distribution (ID) energy scores \(\{\tau_{C^i}\}_{i=1}^{M}\) from previously indexed corpora from the router (i.e., the cosine classifier in a MixLoRA layer, bottom-half of Figure~\ref{fig:mixloradsi}). Before indexing a new corpus, we iterate over its training pseudo-q ueries (i.e., the document representation) to detect OOD cases.  
Since energy scores are higher for OOD data, a token \(c_i\) is flagged as OOD if its energy score exceeds the corresponding threshold \(\tau_{C^i}\) (gray \(c_i\) in Figure~\ref{fig:mixloradsi}). A query is considered OOD if it contains at least one OOD token (Out-of-domain doc, marked with red X marks). If the number of OOD queries for a MixLoRA layer exceeds a predefined threshold \(\delta\), we trigger expert expansion for that layer (dashed LoRA and dashed cosine classifier weight in the MixLoRA layer, bottom-half of Figure~\ref{fig:mixloradsi}).

\paragraph{Improving the router of MixLoRA.}
\label{par:improved_router}
The original MoE design incorporates an auxiliary load balancing loss (Eq.~\ref{eq:load_balance_loss}) to evenly distribute tokens across experts. However, we observe recency bias in the router, where tokens are predominantly routed to newly added experts while ignoring older ones. This issue is exacerbated by the softmax competition in the router, where the unfrozen logits of new router weights dominate over frozen ones, skewing expert selection. 
To address this, we propose replacing the original softmax router, which has unnormalized router weights and takes unnormalized inputs, with a cosine classifier~\cite{gidaris_cosineclassifier_cvpr18}. The cosine classifier router is optimized via cosine embedding loss to better align layer-wise token hidden representations with router weights, i.e., the first term of Eq.~(\ref{eq:router_aux_loss}). Additionally, to prevent router weights from becoming overly similar, we introduce another loss term that encourages new router weights to remain distinct from previous ones, i.e., the second term of Eq.~(\ref{eq:router_aux_loss}). The auxiliary router loss for a MixLoRA layer with \(N\) experts is defined as:
\begin{equation}
\label{eq:router_aux_loss}
\begin{gathered}
    \mathcal{L}_{\text{aux}}(h_{id}; R)=\frac{1}{M}\sum_{i=1}^{M} (1 - \cos(R_N, h_{c_i})) \\ 
    + \sum_{j=1}^{N-1} \max(0, \cos(R_j, R_N)),
\end{gathered}
\end{equation}
where \(\cos(\cdot, \cdot)\) is the cosine function, \(h_{id}=\{h_{c_1}, {\ldots}, h_{c_M}\}\) denotes the docids's hidden representations obtained from the previous Transformer decoder block. Each \(h_{c_i} \in h_{id}\) and the column vectors of \(R\) are unit \(L_2\)-norm weight vectors.

\subsection{RQ-based docids CL strategies}
\label{subsec:rq_cl_strategies}

\paragraph{RQ-based docids.}
\label{subsec:rq_docids}
\citet{mehta_dsipp_emnlp23,huynh_promptdsi_arxiv24} adopt atomic docids due to their robustness against forgetting compared to semantic docids. However, they scale poorly due to the required large softmax space. In this work, we challenge this observation and adopt scalable RQ-based docids. 
Following~\citet{zeng_ripor_www24}, we train the document embeddings with \(M\) RQ codebooks \(C^m\coloneqq\{c_k\}_{k=1}^{K} \in \mathbb{R}^{\text{dim}\times K}\), each containing \(K\) centroids, 
to generate an RQ code of length \(M\) for each document.
The learned RQ centroids are concatenated to the DSI's vocabulary weights \(W_{\text{vocab}}\)
for subsequent optimization: \(W_{\text{RQ}}=\{W_{\text{vocab}}; C^{1}; \ldots; C^{M}\}\),
where \(\{\cdot;\cdot\}\) is concatenation.
Please refer to Appendix~\ref{appendix:rq_docids} for more details about RQ-based docids.

\paragraph{RQ-based docids mask.}
\label{par:rq_docids_mask}
RQ-based docids have a structured design, where each token \(c_i\) is predicted from a specific segment of the extended vocabulary \(W_\text{RQ}\), corresponding to its RQ codebook. To avoid requiring DSI to implicitly learn this structure, we apply a masking strategy that explicitly restricts the softmax computation to the relevant segment of the vocabulary at each decoding step, preventing unnecessary competition between logits. Specifically, for each token \(c_i\) at position \(i\) in the sequence, we define a binary mask \(\mathbf{m}_i\) with the same shape as the output vocabulary, where:  
\begin{equation}
    \label{eq:rq_mask}
    \mathbf{m}_i[j]=
    \begin{cases} 
        1 & \text{if } j \in C^i, \\
        0 & \text{otherwise.}
    \end{cases}
\end{equation}
Let \(\mathbf{z}_i \in \mathbb{R}^{|W_{RQ}|}\) represent the logits produced at decoding step \(i\). During each decoding step, we apply the corresponding mask \(\mathbf{m}_i\) to invalidate irrelevant indices by replacing them with \(-\infty\):
\begin{equation}
    \label{eq:masked_logits}
    \mathbf{z}_i^\prime[j]=
    \begin{cases} 
        \mathbf{z}_i[j]  & \text{if } \mathbf{m}_i[j]=1, \\
        -\infty & \text{if } \mathbf{m}_i[j]=0.
    \end{cases}
\end{equation}
Let us consider the docid \(1{-}2\), which has length \(M=2\), and assume that there are \(K=4\) centroids per codebook. In the first decoding step, we mask the logits corresponding to the last 4 centroids in \(W_{RQ}\in \mathbb{R}^{2\times 4}\) with ${-}\infty$ then apply softmax to generate the first code in the docid \(1\). Similarly, in the next decoding step, we mask the logits corresponding to the first 4 centroids in \(W_{RQ}\in \mathbb{R}^{2\times 4}\) with ${-}\infty$ then apply softmax to generate the second code in the docid \(2\).

\paragraph{Learning compatible RQ-based docids embeddings.}
\label{par:comptatible_rq_docids_embeddings}

To enhance the CL performance of \ourmethod, we employ the slow-learner strategy~\cite{zhang_slca_iccv23} by scaling down the gradient updates of the output vocabulary \(W_\text{RQ}\), while keeping the original portion \(W_\text{vocab}\) frozen. Since we have access to the previous model parameters \(\theta_{t-1}\) before indexing a new corpus, we regularize the updates to the RQ embeddings using KL divergence, loosely aligning the posterior predictive distribution of the current model with that of the previous model at each decoding step:
\begin{equation}
    \label{eq:kl_loss}
    \begin{gathered}
    \mathcal{L}_{\text{KL}}(q, id; \theta_t, \theta_{t-1})= \\
    \frac{1}{M}\sum\limits_{i=1}^{M}\kldiv{P(c_i | c_{<i}, q; \theta_{t-1})}{P(c_i | c_{<i}, q; \theta_{t})}.
    \end{gathered}
\end{equation}
\subsection{Optimization Objective}
The optimization objective for \ourmethod is:
\begin{equation}
    \label{eq:final_loss}
    \begin{gathered}
    \mathcal{L}(Q_{D_t}^{\text{retrieval}}; \theta_t, \theta_{t-1}) =
    \sum_{i=1}^{|Q_{D_t}^{\text{retrieval}}|} \mathcal{L}_{\text{CE}}(q_i, id_{q_i}; \theta_t) \\
    {+}\alpha_1\sum_{l \in L} \mathcal{L}_{\text{aux}}(h^{l-1}, R^{l}){+}\alpha_2\mathcal{L}_{\text{KL}}(q_i, id_{q_i}; \theta_t, \theta_{t-1}),
    \end{gathered}
\end{equation}
where \(\mathcal{L}_{\text{CE}}\) is the log-softmax cross-entropy loss for seq2seq learning, \(L\) denotes the set of MixLoRA layers in DSI's decoder, \(h^{l-1}\) denotes the previous decoder block output, \(R^{l}\) denotes the router of \(l\)-th MixLoRA layer, and \(\alpha_1, \alpha_2\) are hyperparameters. During the pre-training stage on \(D_0\), we do not employ \(\mathcal{L}_\text{KL}\). During continual indexing, if there is expansion, we optimize only the expanded router weights, new LoRAs, and the extended RQ docids vocabulary weights; otherwise, we only optimize the extended RQ docids vocabulary weights (Section~\ref{subsec:rq_docids}). All other parameters are frozen. 
\section{Experiments}
\label{sec:experiments}

\begin{table*}[t]
\setlength{\tabcolsep}{2.5pt}
\renewcommand{\arraystretch}{1.0}
\small
\centering
\scalebox{0.9}{
\begin{tabular}{l ccccccccccc}

\toprule

\multicolumn{11}{c}{\textbf{NQ320k (Recall{@}10/MRR{@}10)}} \\

\textbf{Method} & Params.\(\downarrow\)& \(\mathbf{M_0}\) & \(\mathbf{D_0}\uparrow\) & \(\mathbf{D_1}\uparrow\) & \(\mathbf{D_2}\uparrow\) & \(\mathbf{D_3}\uparrow\) & \(\mathbf{D_4}\uparrow\) & \(\text{AP}_4\uparrow\) & \(\text{BWT}_4\downarrow\) & \(\text{FWT}_4\uparrow\) \\ 

\midrule

\multicolumn{11}{l}{\textbf{Traditional IR models (For reference)}} \\

BM25 & - & 61.3/42.4 & 60.5/41.4 & 54.1/34.3 & 60.3/40.2 & 79.0/{48.4} & 67.1/47.4 & 64.2/42.3 & 1.2/0.4 & 66.1/42.7 \\
DPR & - & 70.4/51.7 & 69.6/50.7 & 68.3/47.0 & 66.2/49.6 & 67.7/48.5 & 70.8/52.9 & 68.5/49.7 & 0.9/0.9 & 68.7/50.1 \\ 

\midrule 

\multicolumn{11}{l}{\textbf{Generative Retrieval (GR) models}} \\

BASE & 235.4\,M & 83.0/70.4 & 18.3/2.5 & 33.3/5.6 & 45.6/11.2 & 67.7/40.8 & 90.6/83.7 & 51.1/28.8 & 47.7/66.3 & 91.0/84.9 \\

\midrule 

\multicolumn{11}{l}{\textbf{Non PEFT-based Continual Learning Generative Retrieval (CLGR) models}} \\

DSI++ & 235.4\,M & 83.0/70.4 & 29.7/5.6 & 41.7/10.1 & {48.5}/17.5 & 45.9/66.1 & {85.6}/{80.3} & 50.3/35.9 & 45.6/55.2 & {87.9}/{82.7} \\

CLEVER(\(n{=}512\)) & 235.4\,M & 83.0/70.4 & 31.0/6.1 & {56.3}/{26.4} & 57.4/30.3 & 74.2/49.8 & \textbf{85.9}/\textbf{81.6} & 61.0/38.9 & 35.0/53.6 & \textbf{90.6}/\textbf{84.8} \\

CLEVER(\(n{=}1024\)) & 235.4\,M & 83.0/70.4 & 40.4/11.1 & \textbf{67.2}/\textbf{46.3} & 62.5/40.2 & 74.2/{58.5} & \textbf{85.9}/\underline{81.1} & 66.1/47.4 & 28.2/42.5 & \underline{90.2}/\underline{84.4} \\ 

\midrule
\multicolumn{11}{l}{\textbf{PEFT-based Continual Learning Generative Retrieval (CLGR) models}} \\

PromptDSI & 4.8\,M &  83.0/70.4 & {59.6}/{44.5} & 28.0/7.9 & 36.0/16.5 & 38.7/27.6 & 51.6/43.0 & 42.8/27.9 & {17.7}/\underline{14.6} & 50.6/32.1 \\ 

CorpusBrain++ & \underline{0.8\,M} &  83.0/70.4 & 6.6/1.9 & 8.7/3.2 & 14.7/8.5 & 9.7/8.9 & 29.6/23.1 & 13.9/9.1 & \textbf{9.9}/\textbf{5.6} & 29.6/23.1 \\ 

Naive Expansion & 1.6\,M & 83.0/70.4 & 48.7/21.0 & 26.3/6.5 & 33.1/12.0 & 43.5/24.6 & 69.3/62.1 & 44.2/25.2 & 38.0/50.0 & 72.5/64.0 \\ 

MixLoRA-DSI\(^{\text{-PT}}\) & 1.2\,M & 83.0/70.4 & \textbf{68.8}/44.0 & 52.4/24.7 & 43.4/20.4 & 46.8/35.4 & 58.1/52.0 & 53.9/35.3 & 14.1/19.3 & 60.8/45.8 \\ 

MixLoRA-DSI\(^{\text{-Expand}}\) & \textbf{0.6\,M} & 80.5/68.1 & \textbf{68.8}/\textbf{51.2} & 52.7/32.6 & 62.5/45.8 & 72.6/63.7 & 75.8/71.5 & 66.5/53.0 & 13.8/17.2 & 76.7/66.3 \\ 

MixLoRA-DSI\(^{\text{-OOD}}\) & 1.6\,M & 80.5/68.1 & 65.1/45.6 & \underline{55.2}/\underline{35.9} & \textbf{72.1}/\textbf{55.6} & \textbf{75.8}/\textbf{70.7} & 75.8/73.3 & \textbf{68.8}/\textbf{56.2} & \underline{12.4}/18.3 & 78.3/71.6 \\

MixLoRA-DSI & 0.9\,M & 80.5/68.1 & \underline{66.1}/\underline{47.2} & 54.1/33.8 & \underline{68.4}/\underline{52.7} & \textbf{75.8}/\underline{69.1} & \underline{76.2}/73.0 & \underline{68.1}/\underline{55.2} & 13.0/18.1 & 78.0/70.0 \\ 

\bottomrule

\end{tabular}
 }
\caption{Models performance after indexing NQ320k's \(D_4\), reported as \{Recall{@}10\}/\{MRR{@}10\}. 
Params. denotes the number of trainable parameters. For PEFT-based CLGR models, we report only the PEFT components, as RQ embeddings dominate the parameter count (i.e., 12.6\,M).
\(\mathbf{M_0}\) shows \(D_0\) performance of the initial checkpoint \(P_{0,0}\). 
\(\mathbf{D_i}\) represents \(P_{4,i}\) on \(D_i\)'s test queries.
\({n}\) denotes the number of samples per corpus used to approximate Fisher Information Matrix in CLEVER. 
Best/Second-best CLGR models are highlighted/underscored, except in \(\mathbf{M_0}\).
}
\label{tab:nq320k_main_result}
\end{table*}
\subsection{Experimental setting}
\label{subsec:experimental_setting}
\paragraph{Datasets.}  
We evaluate on Natural Questions (NQ320k)~\cite{kwiatkowski_nq320k_tacl19} and MSMARCO Passage (MSMARCO)~\cite{bajaj_msmarco_arxiv16}. NQ320k is a small-scale GR benchmark~\cite{sun_genret_neurips23, kishore_incdsi_icml23} with 320k query-document pairs from 108k documents, while MSMARCO is a large-scale IR benchmark with 8.8M passages and 503k queries, evaluated on its 6.9k-query development set.
Following prior works~\cite{kishore_incdsi_icml23,huynh_promptdsi_arxiv24}, we simulate dynamic corpora by splitting each dataset into an initial corpus \(D_0\) (90\% of documents) and four incremental corpora \(D_1\)-\(D_4\) (each 2.5\%). Test queries are partitioned accordingly. Further details are in Appendix~\ref{appendix:datasets}.

\paragraph{Metrics.}
Following~\citet{zeng_ripor_www24,zeng_pag_sigir24}, we use Recall{@}10 (R{@}10) and MRR{@}10 (M{@}10) to evaluate the retrieval performance on both datasets. Following previous works~\cite{mehta_dsipp_emnlp23,huynh_promptdsi_arxiv24}, we adopt Average Performance (AP); Backward Transfer (BWT, aka Forgetting) to measure the effect of indexing current corpus on previous indexed corpora; and Forward Transfer (FWT, aka Learning Performance) to measures the model’s ability to index a new corpus.
Let \(P_{t,i}\) denote the retrieval performance of the model on corpus \(D_i\) after indexing corpus \(D_t\).
With \(0 {\leq} i {<} t\), the CL metrics are defined as follows:
\begin{equation*}
\begin{gathered}
\text{AP}_t = \frac{1}{t}\sum_{i=0}^{t}P_{t,i}; \quad \text{FWT}_t = \frac{1}{t}\sum_{i=1}^{t}P_{i,i};\\
\text{BWT}_t = \frac{1}{t-1}\sum_{i=1}^{t-1}\max_{i^{'}\in\{0,\ldots,t-1\}} (P_{i^{'},i} - P_{t,i}).
\end{gathered}
\end{equation*}
\paragraph{Baselines and model variants.}
\begin{itemize}[topsep=2pt,parsep=2pt,partopsep=2pt,leftmargin=8pt]
    \item \textbf{Traditional IR models}: (1) \textbf{BM25}~\cite{robertson_bm25_now09} is a strong sparse retrieval baseline. (2) \textbf{DPR}~\cite{karpukhin_dpr_emnlp20} is a popular BERT-based dual-encoder.
    
    \item \textbf{GR models}: \textbf{BASE} uses RQ-based docids~\cite{zeng_ripor_www24} and is sequentially fine-tuned on new corpora. This setup is equivalent to RQ-based docid \textbf{Ultron}~\cite{zhou_ultron_arxiv22}.
    
    \item \textbf{Continual Learning Generative Retrieval (CLGR) models}: 
    we compare CLGR models in terms of the non-rehearsal CL strategies which starts from a T5 checkpoint pre-trained on \(D_0\).
    (1) \textbf{DSI++}~\cite{mehta_dsipp_emnlp23} continually fine-tunes DSI with Sharpness-Aware Minimization~\cite{foret_sam_2020}. 
    (2) \textbf{CLEVER}~\cite{chen_clever_cikm23} 
    continually fine-tunes DSI with Elastic Weight Consolidation regularization (EWC)~\cite{kirkpatrick_ewc_pnas17}, closely aligned with the CLEVER$^{\text{-MLE(d-)}}$ variant from the original work.
    (3) \textbf{PromptDSI}~\cite{huynh_promptdsi_arxiv24}  uses prefix-tuning~\cite{li_prefixtuning_21} and neural topic embeddings. 
    (4) \textbf{CorpusBrain++}~\cite{guo_corpusbrain++_arxiv24} continually fine-tunes a set of adapters~\cite{houlsby_adapter_icml19} to adapt to new documents.
    (5) \textbf{Naive Expansion}: replaces selected T5’s FFNs with 2-expert MixLoRAs, expanding by one LoRA per layer for each new corpus; uses the original MoE router and load balancing loss. 

    \item \textbf{MixLoRA-DSI}: starts from a 2-expert MixLoRA architecture checkpoint pre-trained on \(D_0\) with the proposed improved router, variants include:
    (1) \textbf{MixLoRA-DSI\(^{\text{-PT}}\)} starts from a normal T5 checkpoint pre-trained on \(D_0\).
    (2) \textbf{MixLoRA-DSI\(^{\text{-Expand}}\)} does not expand any further during continual indexing.
    (3) \textbf{MixLoRA-DSI\(^{\text{-OOD}}\)} does not employ energy-based dynamic expansion.
\end{itemize}
Further implementation details are in Appendix~\ref{appendix:implementation_details}.

\begin{table*}[t]
\setlength{\tabcolsep}{2.5pt}
\renewcommand{\arraystretch}{1.0}
\small
\centering
\scalebox{0.9}{
\begin{tabular}{lccccccccccc}

\toprule

\multicolumn{11}{c}{\textbf{MSMARCO (Recall@10/MRR@10)}} \\

\textbf{Method} & Params.\(\downarrow\) & \(\mathbf{M_0}\) & \(\mathbf{D_0}\uparrow\) & \(\mathbf{D_1}\uparrow\) & \(\mathbf{D_2}\uparrow\) & \(\mathbf{D_3}\uparrow\) & \(\mathbf{D_4}\uparrow\) & \(\text{AP}_4\uparrow\) & \(\text{BWT}_4\downarrow\) & \(\text{FWT}_4\uparrow\) \\ 

\midrule

\multicolumn{11}{l}{\textbf{Traditional IR models (For reference)}} \\

BM25 & - & 39.8/19.1 & 38.4/18.2 & 39.1/19.0 & 38.8/19.4 & 32.8/13.5 & 40.3/19.3 & 37.9/17.9 & 1.0/0.5 & 38.4/18.1 \\
DPR & - & 53.2/26.0 & 51.6/24.6 & 49.5/26.2 & 54.1/25.4 & 48.3/22.3 & 49.4/24.5 & 50.6/24.6 & 0.5/0.7 & 50.4/24.9 \\

\midrule 

\multicolumn{11}{l}{\textbf{Generative Retrieval models}} \\

BASE & 235.4\,M & 34.3/17.7 & 7.5/1.4 & 8.2/2.5 & 7.7/1.2 & 22.4/5.9 & 87.7/69.1 & 26.7/16.0 & 53.6/55.2 & 78.3/70.8 \\

\midrule 

\multicolumn{11}{l}{\textbf{Non PEFT-based Continual Learning Generative Retrieval (CLGR) models}} \\

DSI++ & 235.4\,M & 34.3/17.7 & 7.6/1.6 & 6.5/1.7 & 6.1/0.8 & 17.8/4.9 & \textbf{90.3}/\textbf{68.0} & 25.7/15.4 & 59.1/50.9 & \textbf{82.6}/\textbf{65.7} \\

CLEVER(n=512) & 235.4\,M & 34.3/17.7 & 14.4/4.1 & 25.5/9.9 & 19.9/7.6 & 27.6/12.6 & \underline{81.2}/\underline{67.6} & 33.7/20.4 & 45.1/43.2 & \underline{78.6}/\underline{64.2} \\

CLEVER(n=1024) & 235.4\,M & 34.3/17.7 & 18.7/6.2 & 	\textbf{41.3}/\textbf{23.0} & \textbf{29.6}/\textbf{13.8} & 31.6/13.0 & 70.1/60.2 & \textbf{38.3}/\textbf{23.2} & 33.1/36.0 & 72.4/60.6 \\

\midrule

\multicolumn{11}{l}{\textbf{PEFT-based Continual Learning Generative Retrieval (CLGR) models}} \\

PromptDSI & 101.5\,M & 34.3/17.7 & 17.6/7.5 & 18.5/8.6 & 12.8/5.1 & 14.4/5.1 & 39.0/34.3 & 20.4/12.1 & {32.0}/{31.3} & 48.9/42.0 \\

Naive Expansion & 1.6\,M & 34.3/17.7 & 24.5/10.4 & 20.7/8.5 & 17.3/6.5 & 24.7/10.7 & 65.6/56.0 & 30.6/18.4 & 28.5/30.8 & 58.2/49.4 \\

CorpusBrain++ & \underline{0.8\,M} & 34.3/17.7 & \textbf{28.9}/\textbf{15.4} & 30.4/\underline{15.0} & 28.6/\underline{11.7} &  28.7/15.6 & 24.3/40.9 & 28.2/19.7 & \textbf{3.8}/\textbf{4.4} & 30.1/24.6 \\

MixLoRA-DSI\(^{\text{-PT}}\) & 1.2\,M & 34.3/17.7 & 25.0/12.1 & 26.1/12.0 & 20.9/7.5 & 27.6/14.2 & 47.4/41.6 & 29.4/17.5 & \underline{14.4}/\underline{18.1} & 42.6/35.5 \\

MixLoRA-DSI\(^{\text{-OOD}}\) & 1.6\,M & 34.0/17.8 & 27.9/\underline{13.6} & 29.9/13.4 & 25.5/10.8 & \underline{33.3}/\underline{17.7} & 58.4/44.5 & 35.0/20.0 & 16.6/19.8 & 51.9/40.3 \\

MixLoRA-DSI & \textbf{0.6\,M} & 34.0/17.8 & \underline{28.1}/\underline{13.6} & \underline{31.0}/13.5 & \underline{29.1}/11.4 & \textbf{34.5}/\textbf{18.2} & 59.1/46.3 & \underline{36.4}/\underline{20.6} & 16.5/20.4 & 53.5/41.7 \\

\bottomrule

\end{tabular}
}
\caption{Models performance after indexing MSMARCO's \(D_4\). Refer to Table~\ref{tab:nq320k_main_result} for a detailed caption.}
\label{tab:msmarco_main_result}
\end{table*}
\subsection{Main Results}
\label{subsec:main_results}

Table~\ref{tab:nq320k_main_result} (NQ320k) and Table~\ref{tab:msmarco_main_result} (MSMARCO) show that even a non-learning baseline like BM25 suffers from performance drift; while zero-shot retrieval with DPR exhibits slight forgetting. Both highlight that forgetting is unavoidable in dynamic corpora.

\paragraph{NQ320k.} 
Both GR and non-PEFT CLGR models fine-tune all parameters, exhibiting strong recency bias~\cite{smith_rehearsalfreecl_cvpr23}, where models overfit recent corpora and suffer catastrophic forgetting on earlier ones.
While CLEVER better preserves performance on new corpora, it incurs substantial memory costs (Section~\ref{par:ablate_memory}) and struggles to retain \(D_0\) performance without rehearsal.

In contrast, PEFT-based CLGR methods prioritize retaining performance on the initial corpus \(D_0\) by freezing the pre-trained backbone, often at the expense of adaptability to new corpora. For example, {CorpusBrain++} achieves the lowest forgetting (lowest BWT$_4$), but lacks plasticity, resulting in poor AP$_4$ and FWT$_4$.
{MixLoRA-DSI\(^\text{-PT}\)} outperforms {Naive Expansion}, {PromptDSI}, and {CorpusBrain++} in terms of AP, demonstrating the effectiveness of MixLoRA layers with improved routers even without pre-training. Despite fine-tuning fewer than 2\,M parameters, MixLoRA-DSI variants exceed {CLEVER} in AP$_4$ while suffering much less from forgetting.
{MixLoRA-DSI\(^\text{-Expand}\)} further highlights the importance of expansion for incorporating novel information. With layer-wise OOD-driven dynamic expansion, MixLoRA-DSI achieves sublinear parameter growth: using just 60\% of MixLoRA-DSI\(^\text{-OOD}\)'s trainable parameters while maintaining over 98\% of AP$_4$ and FWT$_4$ and better preserve performance on \(D_0\). 

\paragraph{MSMARCO.}  
Scaling to MSMARCO presents greater challenges due to its massive size. However, the overall trend mirrors NQ320k: non-PEFT methods suffer from severe forgetting, with CLEVER reducing this at the cost of substantial memory overhead.
Interestingly, CorpusBrain++ demonstrates significant robustness to forgetting. We hypothesize this is due to its continual fine-tuning of adapters with massive amounts of new data, which contrasts with MixLoRA-DSI’s freeze-and-expand strategy.
Our dynamic expansion strategy remains effective at controlling parameter growth. For MSMARCO, MixLoRA-DSI coincides with the non-expansion variant, outperforms PromptDSI and MixLoRA-DSI\(^\text{-OOD}\) in terms of AP$_4$, ranking second only to CLEVER(\(n {=} 1024\)), while exhibiting significantly lower forgetting. We attribute this gap to full-model updates providing greater capacity for memorizing large corpora.

\subsection{Discussions}
\label{subsec:discussions}
\begin{table}[t]
\setlength{\tabcolsep}{2.5pt}
\centering
\scalebox{0.95}{
\begin{tabular}{l|cccc}

\toprule

Method & EWC & Train & Inference & Storage \\

\midrule

DSI++              & -    & 10.2 & 21.1 & \textbf{0.9} \\
CLEVER(\(n\)=512)  & 18.4 & 12.3 & 21.1 & 6.2 \\
CLEVER(\(n\)=1024) & 32.4 & 12.3 & 21.1 & 6.2 \\
PromptDSI          & -    & \textbf{6.3} & 18.4 & \textbf{0.9} \\
CorpusBrain++      & -    & \textbf{6.3} & 30.9 & \textbf{0.9} \\
\midrule
MixLoRA-DSI        & -    & 8.1 & \textbf{18.0} & \textbf{0.9} \\

\bottomrule

\end{tabular}
}
\caption{Memory footprints (GiB) during training and inference measured with a batch size of 128; and storage. CLEVER requires significant GPU memory to approximate Fisher Information Matrix due to EWC, and demands more storage.}
\label{tab:ablate_memory}
\end{table}
\paragraph{Memory comparison.}
\label{par:ablate_memory}
In Table~\ref{tab:ablate_memory}, we compare the memory usage of CLGR models. CLEVER incurs high memory costs in both storage and GPU usage due to EWC regularization, which scales with \(n\) (i.e., the number of prior samples used to estimate parameter importance via log-likelihood gradients). While a larger \(n\) helps mitigate forgetting, it significantly increases memory overhead. CLEVER must also store previous model checkpoints and their corresponding gradients.
CorpusBrain++ is efficient during training, but its adapters require high memory during inference.
In contrast, MixLoRA-DSI is slightly more memory-intensive during training but significantly more efficient during inference. It requires saving only the latest model and a temporary copy during training to compute \(\mathcal{L}_{\text{KL}}\) (Eq.~\ref{eq:kl_loss}).

\paragraph{Ablation studies.}  
\begin{table}[t]
\scalebox{0.85}{
\setlength{\tabcolsep}{1.2pt}
\centering
\begin{tabular}{ccccc|ccccc}

\toprule
\multicolumn{5}{c|}{Methods} & \multicolumn{2}{c}{\(\text{AP}_4 \uparrow\)} & \multicolumn{2}{c}{\(\text{BWT}_4 \downarrow\)} \\
{Mask.} & {CL.} & {Rout.} & {PT.} & {OOD.} & R@10 & M@10 & R@10 & M@10 \\

\midrule 
\xmark & \xmark & \xmark & \xmark & \xmark & 36.2 & 20.7 & 33.1 & 38.7 \\
\cmark & \xmark & \xmark & \xmark & \xmark & 44.2 & 25.2 & 35.8 & 49.7 \\
\cmark & \cmark & \xmark & \xmark & \xmark & 52.5 & 32.8 & 15.8 & 21.1 \\ 
\cmark & \cmark & \cmark & \xmark & \xmark & 53.2 & 34.6 & \underline{12.5} & \textbf{15.3} \\
\cmark & \cmark & \cmark & \cmark & \xmark & \textbf{68.8} & \textbf{56.2} & \textbf{12.4} & 18.3 \\
\cmark & \cmark & \cmark & \cmark & \cmark & \underline{68.1} & \underline{55.2} & 13.0 & \underline{18.1} \\

\bottomrule

\end{tabular}
}
\caption{Ablation results on NQ320k for MixLoRA-DSI's design: RQ-based docid mask (Mask.), RQ-based docid CL strategies (CL.), Improved router (Rout.), Pre-training on \(D_0\) (PT.), and OOD-driven dynamic expansion (OOD.). Please refer to Appendix~\ref{appendix:ablation_msmarco} for ablation results on MSMARCO.
}
\label{tab:ablation_nq320k}
\end{table}
Table~\ref{tab:ablation_nq320k} shows ablations of MixLoRA-DSI. The RQ-based docid mask provides a strong performance boost in \(\text{AP}_4\) (+8.0 Recall{@}10) but does not mitigate forgetting. Incorporating RQ-based docid CL strategies further enhances retrieval (+8.4 Recall{@}10) while significantly reducing forgetting (-20.0 Recall{@}10). The improved router further increases overall performance. Pre-training on \(D_0\) has the most substantial impact on retrieval (+15.6 Recall{@}10), underscoring the importance of a strong initialization. OOD-driven expansion limits unnecessary growth while preserving performance. Overall, all components contribute to the strong stability-plasticity trade-off of MixLoRA-DSI.

\paragraph{Token routing analysis.}
\begin{figure}[t]
      \centering     
      \includegraphics[width=0.3\linewidth]{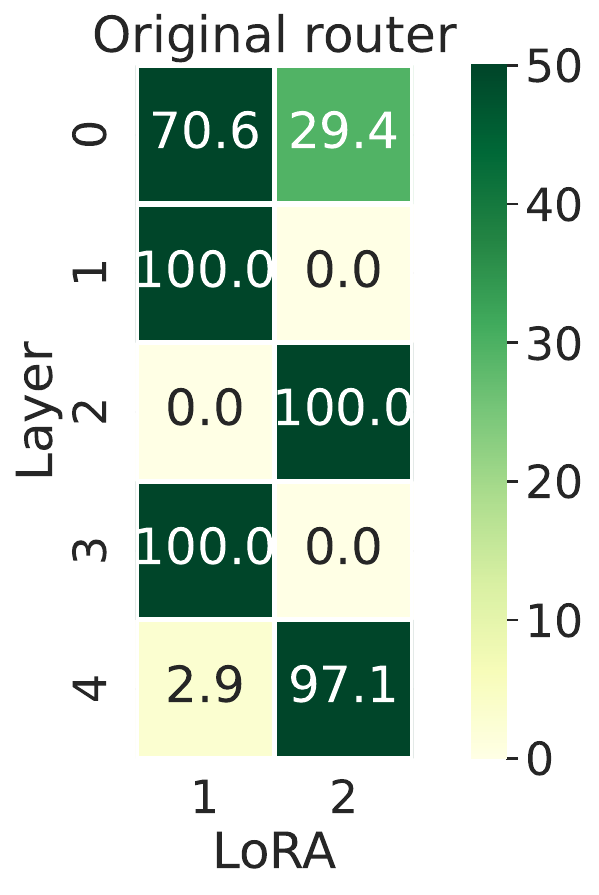}
      \includegraphics[width=0.3\linewidth]{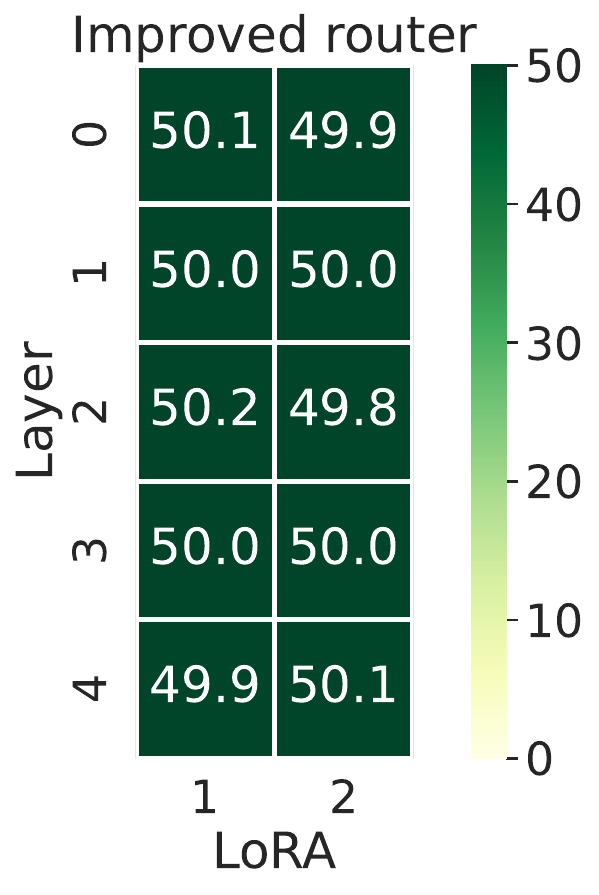} \\
      \includegraphics[width=0.48\linewidth]{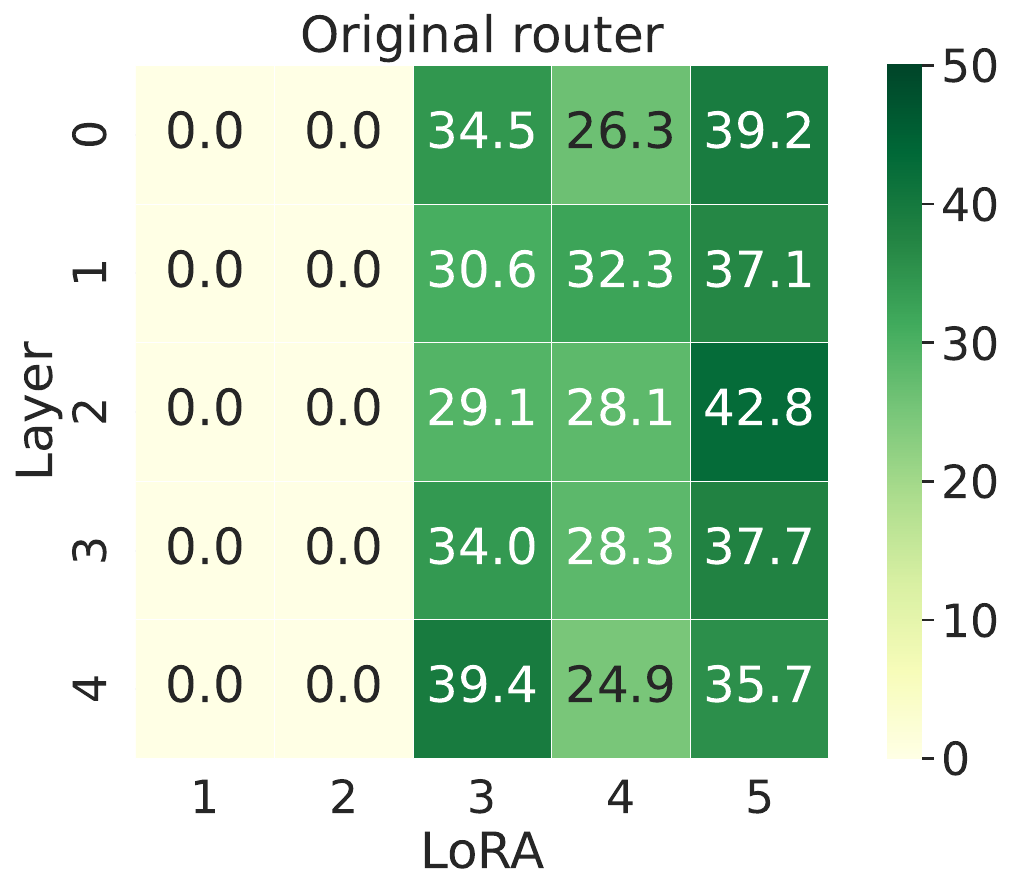}
      \includegraphics[width=0.48\linewidth]{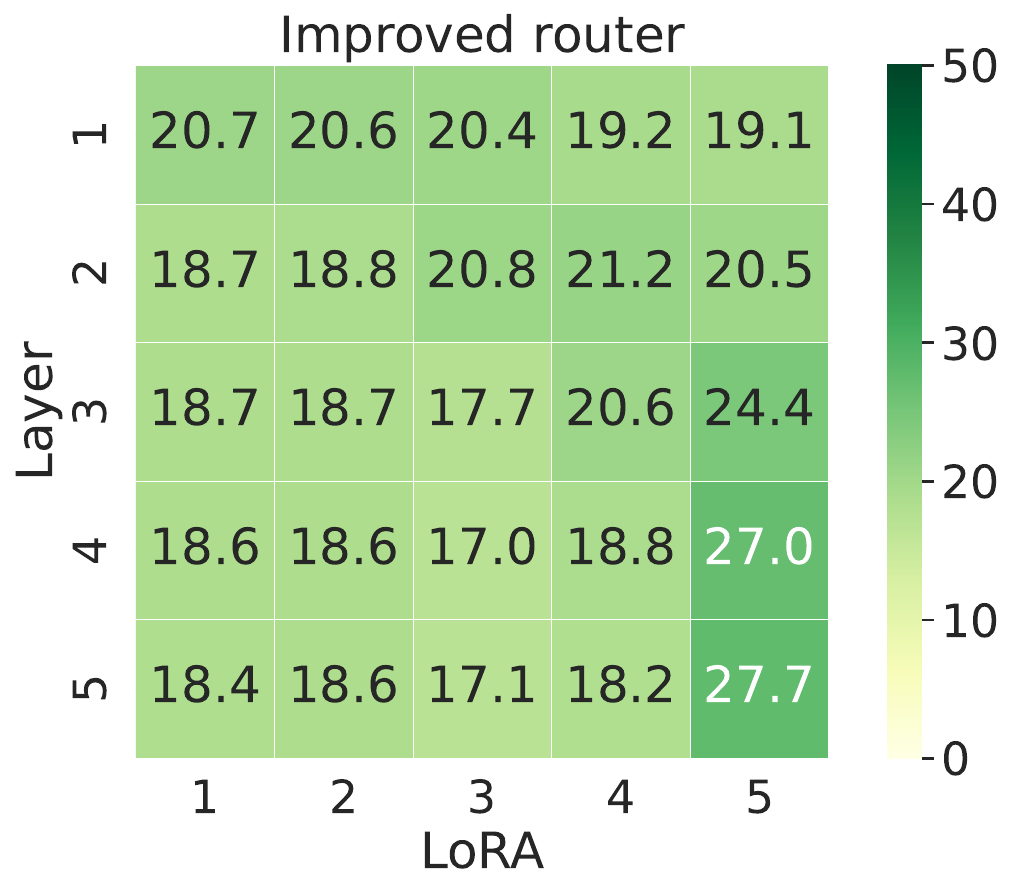}
      \caption{Token routing analysis of the original MoE routers (left) and our proposed improved routers (right) when training on \(D_1\) (Top row) and \(D_4\) (Bottom row). Each cell indicates the percentage of tokens being routed to a LoRA expert in a layer. Our improved routers achieve balanced routing while the original router suffers from recency bias.}
      \label{fig:token-routing}
\end{figure}
Figure~\ref{fig:token-routing} analyzes the routing behavior of the original and improved routers. Naive Expansion starts with 2-expert MixLoRAs on \(D_1\) and linearly adds new LoRAs per corpus. However, its softmax router, with unnormalized weights and inputs, induces competition, causing new tokens to be predominantly assigned to recently added experts. In contrast, our improved router, guided by a novel auxiliary loss, achieves a more balanced routing distribution while still prioritizing newly added LoRAs (e.g., the 5th for \(D_4\)). This demonstrates that our approach effectively preserves prior experts' knowledge while enabling new LoRAs to capture novel information.  

\paragraph{Layer-wise OOD queries analysis.} 
\begin{table}[t]
\setlength{\tabcolsep}{2.5pt}
\renewcommand{\arraystretch}{1.0}
\centering
\scalebox{0.95}{
\begin{tabular}{c|cccc|cccc}

\toprule
\multirow{2}{*}{Layer} & \multicolumn{4}{c|}{MixLoRA-DSI} & \multicolumn{4}{c}{MixLoRA-DSI\(^{\text{-PT}}\)} \\
& \(D_1\) & \(D_2\) & \(D_3\) & \(D_4\) & \(D_1\) & \(D_2\) & \(D_3\) & \(D_4\) \\

\midrule 

1 & 3.6 & 0.0 & 0.0 & 0.0 & - & 81.2 & 55.8 & 22.0 \\
2 & 1.6 & 0.0 & 0.0 & 0.0 & - & 43.1 & 19.3 & 5.6 \\
3 & 1.2 & 0.0 & 0.0 & 0.0 & - & 33.0 & 6.9 & 7.1 \\
4 & 1.3 & 0.0 & 0.0 & 0.0 & - & 35.8 & 5.9 & 5.3 \\
5 & 2.0 & 0.0 & 0.0 & 0.0 & - & 47.2 & 14.6 & 2.8 \\
\midrule

\# LoRA & 10 & 15 & 15 & 15 & 10 & 15 & 18 & 19 \\

\bottomrule

\end{tabular}
}
\caption{Percentage of layer-wise OOD queries for MixLoRA-DSI and MixLoRA-DSI\(^{\text{-PT}}\) from Table~\ref{tab:nq320k_main_result} before training on each new corpus of NQ320k. 
`-' denotes cases where OOD detection are inapplicable.
}
\label{tab:ood_mixlora_dsi}
\end{table} 
Table~\ref{tab:ood_mixlora_dsi} presents the percentage of OOD queries. In the left-half, MixLoRA-DSI, with five 2-expert MixLoRA layers in the decoder, expands only once before \(D_1\), resulting in total 15 LoRAs, as pre-training on \(D_0\) provides broad general knowledge.
In contrast, MixLoRA-DSI\(^\text{-PT}\) (right-half) starts from a pre-trained T5 checkpoint, initializing all LoRAs and routers from scratch with no prior energy scores for OOD detection. Consequently, OOD estimation begins only after indexing \(D_1\). As indexing progresses, energy scores stabilize, reducing OOD detections and slowing expansion, ensuring sublinear parameter growth.

\section{Related Work}
\label{sec:related_work}
\paragraph{Generative Retrieval (GR).}  
Docids remain a core research focus in GR~\cite{bevilacqua_seal_neurips22,li_minder_acl23,sun_genret_neurips23,zhang_tsgen_sigir24,zeng_pag_sigir24,liu_asi++_arxiv24}. 
Recent works improve GR through learning-to-rank~\cite{li_ltrgr_aaai24,zeng_ripor_www24} and multi-graded retrieval~\cite{tang_gr2_neurips24}; however, these methods are not designed for dynamic corpora. 
GR over dynamic corpora is gaining traction~\cite{mehta_dsipp_emnlp23,chen_clever_cikm23,guo_corpusbrain++_arxiv24}, 
with recent efforts addressing rehearsal-free CL~\cite{huynh_promptdsi_arxiv24} and temporal IR~\cite{kim_dynamicgenir_emnlp24}.
We note that related works such as IncDSI~\cite{kishore_incdsi_icml23}, which differs from autoregressive GR, and L$^2$R~\cite{cai_l2r_cikm23}, a rehearsal-based dense retrieval method, fall outside the scope of this work.
MixLoRA-DSI pushes dynamic corpora GR towards task-agnostic, rehearsal-free, and scalable retrieval by leveraging RQ-based docids~\cite{zeng_ripor_www24}, introducing PEFT, and employing a dynamic expansion strategy. 
This challenges prior works that primarily adopt atomic docids and evaluate on small-scale datasets.

\paragraph{Parameter-efficient Fine-tuning (PEFT).}  
PEFT introduces lightweight trainable components to efficiently train large models. The three major methods are Adapters~\cite{houlsby_adapter_icml19}, Prompts~\cite{lester_ptuning_21,li_prefixtuning_21}, and LoRA~\cite{hu_lora_iclr22}. LoRA is recently integrated into  MoE~\cite{shazeer_sparsemoe_iclr17,fedus_switchtransformer_jmlr22}, results in MixLoRA and its variants in vision and NLP~\cite{dou_loramoe_acl24,yu_moeloraclvision_cvpr24,wu_mole_iclr24}. However, its application in IR remains unexplored. We pioneer its use in dynamic corpora GR, introducing IR-specific enhancements: an improved router with novel auxiliary loss and an energy-based dynamic expansion strategy to address the challenges of dynamic corpora.

\section{Conclusion}
\label{sec:conclusion}
We introduce MixLoRA-DSI, the first dynamically expandable framework for rehearsal-free GR over dynamic corpora, which is task-agnostic, rehearsal-free, and scalable. Our layer-wise OOD-driven expansion strategy ensures sublinear parameter growth, and the proposed cosine classifier with novel auxiliary loss improves expert specialization and balance routing. Integrating with CL strategies for RQ-based docids, MixLoRA-DSI outperforms prior GR methods in efficiency and robustness.

\section*{Limitations}
Our work does not incorporate recent ranking optimization techniques~\cite{li_ltrgr_aaai24,zeng_ripor_www24}, as state-of-the-art RQ-based docid methods often require multi-step training~\cite{zeng_ripor_www24,zeng_pag_sigir24}, which may exacerbate forgetting in dynamic corpora. Consequently, all GR models lag behind traditional IR on MSMARCO in this realistic setting.
Due to resource constraints, we experiment only with T5-base. Although prior work~\cite{pradeep_genirscaling_emnlp23} suggests that GR performance does not scale linearly with model size, exploring larger models is left for future work. We also do not perform extensive hyperparameter tuning, yet observe consistent gains from each component.
While we adopt a freeze-and-expand strategy, CorpusBrain++ shows that continually fine-tuning a fixed set of adapters yields strong robustness to forgetting. This raises an open question about the trade-off between dynamic expansion and parameter budget.
We further acknowledge the need to assess generalization across diverse retrieval tasks (e.g., KILT~\cite{petroni_kilt_naacl21}) but defer this to future work due to space constraints.
Finally, GR is not suitable for zero-shot retrieval, as indexing necessitates model updates.
Addressing these challenges offers promising directions for future work.

\section*{Acknowledgments}
This material is based on research sponsored by Defense Advanced Research Projects Agency (DARPA) under agreement number HR0011-22-2-0047. The U.S. Government is authorized to reproduce and distribute reprints for Governmental purposes notwithstanding any copyright notation thereon. The views and conclusions contained herein are those of the authors and should not be interpreted as necessarily representing the official policies or endorsements, either expressed or implied, of DARPA or the U.S. Government.
This work is supported by the Australian Research Council Discovery Early Career Researcher Award DE250100032 
and Monash eResearch capabilities, including M3.
Trung Le was partly supported by the Air Force Office of Scientific Research under award number FA2386-23-1-4044.

\bibliography{custom}

\appendix
\section{Datasets}
\label{appendix:datasets}

The detailed statistics of NQ320k~\cite{kwiatkowski_nq320k_tacl19} and MSMARCO~\cite{bajaj_msmarco_arxiv16} are shown in Table~\ref{tab:nq320k-dataset} and~\ref{tab:msmarco-dataset}, respectively. For reproducibility, we adopt pseudo-queries provided by previous works: MSMARCO uses 10 per document~\cite{zeng_ripor_www24}, and NQ320k up to 15~\cite{kishore_incdsi_icml23}.

It is worth noting that MSMARCO contains 28 times more documents but only 2 times more annotated queries than NQ320k. To prevent data leakage in the dynamic corpora setting, we ensure that the splits ($D_0$-$D_4$) do not contain information from subsequent corpora, which results in fewer annotated queries, further impacting the quality of RQ-based docids construction.

\begin{table*}[ht]
\small
\centering
\begin{tabular}{ccccc}
\toprule
\multicolumn{5}{c}{NQ320k}\\ \midrule
Split & Document & Annotated Queries & Train Pseudo Queries & Test Queries \\ \midrule
$D_0$ & 9,8743 & 276,493 & 1,480,538 & 6,998 \\
$D_1$ & 2,743 & - &  41,132 & 357 \\
$D_2$ & 2,743 & - & 41,136 & 136 \\
$D_3$ & 2,743 & - & 41,138 & 62 \\
$D_4$ & 2,743 & - & 41,141 & 277 \\
\bottomrule
\end{tabular}
\caption{The NQ320k dataset statistics used in our study.}
\label{tab:nq320k-dataset}
\end{table*}

\begin{table*}[ht]
\small
\centering
\begin{tabular}{ccccc}
\toprule
\multicolumn{5}{c}{MSMARCO}\\ \midrule
Split & Document & Annotated Queries & Train Pseudo Queries & Test Queries \\ \midrule
$D_0$ & 7,957,640 & 452,746 & 79,576,202  &  6,247 \\
$D_1$ & 221,045 & - &  2,210,445  &  184  \\
$D_2$ & 221,045 & - &  2,210,445  &  196  \\
$D_3$ & 221,045 & - &  2,210,447  &  174  \\
$D_4$ & 221,048 & - &  2,210,471  &  154  \\
\bottomrule
\end{tabular}
\caption{The MSMARCO dataset statistics used in our study.}
\label{tab:msmarco-dataset}
\end{table*}

\begin{table*}[ht]
\small
\centering
\begin{tabular}{ccccc}
\toprule
\multicolumn{5}{c}{LongEval}\\ \midrule
Split & Document & Annotated Queries & Train Pseudo Queries & Test Queries \\ \midrule
$D_0$ (2022-10)	& 2,137,584	& 9,005 & 8,790,469 & 1,003 \\
$D_1$ (2022-11)	& 49,628 & - & 248,140 & 1,127 \\ 
$D_2$ (2022-12)	& 468,940 & - & 2,344,700 & 1,128 \\
$D_3$ (2023-01)	& 35,777 & - & 178,885 & 1,163 \\
\bottomrule
\end{tabular}
\caption{The LongEval dataset statistics used in our study.}
\label{tab:longeval-dataset}
\end{table*}
\section{Additional results}
\label{appendix:additional_results}
\subsection{Ablation on the choice of using RQ-based document identifiers}
\begin{table*}[t]
\setlength{\tabcolsep}{2.5pt}
\renewcommand{\arraystretch}{1.0}
\centering
\scalebox{0.9}{
\begin{tabular}{l|ccccc}

\toprule
\multirow{2}{*}{Docid variant} & \multicolumn{2}{c}{\(\text{AP} \uparrow\)} & \multicolumn{2}{c}{\(\text{BWT} \downarrow\)} & Extended vocab size \(\downarrow\) \\
& Recall@10 & MRR@10 & Recall@10 & MRR@10 & MiB \\

\midrule 
Atomic & 44.4 & 24.1 & \textbf{0.0} & \textbf{0.0} & 318 \\
PQ-based & 3.4 & 1.6 & 1.0 & 1.5 & \textbf{48} \\
RQ-based & \textbf{53.2} & \textbf{34.6} & 12.6 & 15.5 & \textbf{48} \\

\bottomrule

\end{tabular}
}
\caption{\(\text{MixLoRA-DSI}^{-\text{PT}}\) with different docid variants on \(D_0\) of NQ320k.}
\label{tab:ablation_rqdocid}
\end{table*}

While MixLoRA-DSI is compatible with any semantic docid (i.e., docids derived from quantization or hierarchical clustering over document embeddings~\cite{zeng_ripor_www24, zeng_pag_sigir24}), generative retrieval generally depends on high-quality docids, which are typically fixed after construction. We provide additional experiments on NQ320k to justify the choice of RQ-based docid in Table~\ref{tab:ablation_rqdocid}.

We explored another popular semantic docid representation, Product Quantization (PQ)-based docids, used in prior work such as CLEVER~\cite{chen_clever_cikm23} and Ultron~\cite{zhou_ultron_arxiv22}. On\(D_0\) , even with exact search via FAISS~\cite{johnson_faiss_tbd19}, PQ-based docids achieved only 13.0\% Recall@10 and 6.9\% MRR@10, compared to 76.6\% Recall@10 and 60.0\% MRR@10 with RQ-based docids. As a result, MixLoRA-DSI-PT with PQ-based docids yields poor overall performance.

Another widely used docid in continual learning for generative retrieval is the atomic docid, as adopted by DSI++~\cite{mehta_dsipp_emnlp23}, IncDSI~\cite{kishore_incdsi_icml23}, and PromptDSI~\cite{huynh_promptdsi_arxiv24}. While atomic docid is less prone to forgetting, they come with a substantial memory cost: the vocabulary embedding matrix grows linearly with the number of documents. For example, in MSMARCO, this would require an approximately 25\,GiB extended vocabulary to store 8.8 million unique document embeddings. This makes atomic docid impractical for large-scale applications, especially given our goal of efficiency and sub-linear parameter growth. In contrast, both PQ- and RQ-based docids use a fixed-size extended vocabulary, which adds minimal memory overhead and still manages a good stability-plasticity trade-off.

\subsection{Ablation studies on MSMARCO}
\label{appendix:ablation_msmarco}
We provide the ablation study on MS MARCO in Table~\ref{tab:ablation_msmarco}. The results demonstrate that each proposed component contributes to the overall stability-plasticity tradeoff, leading to the improved performance of MixLoRA-DSI, similar to the findings on NQ320k in Table~\ref{tab:ablation_nq320k}.

\begin{table}[t]
\scalebox{0.85}{
\setlength{\tabcolsep}{1.2pt}
\renewcommand{\arraystretch}{1.0}
\centering
\begin{tabular}{ccccc|ccccc}

\toprule
\multicolumn{5}{c|}{Methods} & \multicolumn{2}{c}{\(\text{AP}_4 \uparrow\)} & \multicolumn{2}{c}{\(\text{BWT}_4 \downarrow\)} \\
{Mask.} & {CL.} & {Rout.} & {PT.} & {OOD.} & R@10 & M@10 & R@10 & M@10 \\

\midrule 
\xmark & \xmark & \xmark & \xmark & \xmark & 29.4 & 17.5 & \textbf{14.4} & \underline{18.1} \\
\cmark & \xmark & \xmark & \xmark & \xmark & 30.6 & 18.4 & 28.5 & 30.8 \\
\cmark & \cmark & \xmark & \xmark & \xmark & 34.3 & 20.0 & 17.3 & 20.1 \\ 
\cmark & \cmark & \cmark & \xmark & \xmark & 34.5 & 19.7 & \underline{15.0} & \textbf{17.9} \\
\cmark & \cmark & \cmark & \cmark & \xmark & \underline{35.7} & \underline{20.2} & 16.9 & 20.3 \\
\cmark & \cmark & \cmark & \cmark & \cmark & \textbf{36.4} & \textbf{20.6} & 16.5 & 20.4 \\

\bottomrule

\end{tabular}
}
\caption{Ablation results on MSMARCO for MixLoRA-DSI's design: RQ-based docid mask (Mask.), RQ-based docid CL strategies (CL.), Improved router (Rout.), Pre-training on \(D_0\) (PT.), and, OOD-driven dynamic expansion (OOD.).
}
\label{tab:ablation_msmarco}
\end{table}

\begin{table}[t]
\centering
\scalebox{0.85}{
\setlength{\tabcolsep}{1.5pt}
\renewcommand{\arraystretch}{1}
\begin{tabular}{l|cc|cc|cc}
\toprule
\textbf{AP} & \multicolumn{2}{c|}{$\alpha_2=0.05$} & \multicolumn{2}{c|}{$\alpha_2=0.1$} & \multicolumn{2}{c}{$\alpha_2=0.5$} \\
\cmidrule(lr){1-1} \cmidrule(lr){2-3} \cmidrule(lr){4-5} \cmidrule(lr){6-7}
Metrics & R@10 & M@10 & R@10 & M@10 & R@10 & M@10 \\
\midrule
$\alpha_1=0.05$ & 65.2 & 50.8 & 66.7 & 52.3 & 59.5 & 40.6 \\
$\alpha_1=0.1$  & 64.9 & 50.2 & 67.0 & 52.3 & 59.7 & 40.8 \\
$\alpha_1=0.5$  & 66.4 & 51.8 & \textbf{68.1} & \textbf{55.2} & 61.9 & 45.3 \\
\bottomrule
\end{tabular}
}
\caption{Average performance (AP) of MixLoRA-DSI on NQ320k under different combinations of $\alpha_1$ and $\alpha_2$.}
\label{tab:alpha_ablation_ap}
\end{table}

\begin{table}[t]
\centering
\scalebox{0.85}{
\setlength{\tabcolsep}{1.5pt}
\renewcommand{\arraystretch}{1}
\begin{tabular}{l|cc|cc|cc}
\toprule
\textbf{BWT} & \multicolumn{2}{c|}{$\alpha_2=0.05$} & \multicolumn{2}{c|}{$\alpha_2=0.1$} & \multicolumn{2}{c}{$\alpha_2=0.5$} \\
\cmidrule(lr){1-1} \cmidrule(lr){2-3} \cmidrule(lr){4-5} \cmidrule(lr){6-7}
Metrics & R@10 & M@10 & R@10 & M@10 & R@10 & M@10 \\
\midrule
$\alpha_1=0.05$ & 15.3 & 21.0 & 13.3 & 20.6 & 12.9 & 15.2 \\
$\alpha_1=0.1$  & 18.3 & 26.2 & 12.8 & 17.7 & \textbf{8.0}  & \textbf{5.0}  \\
$\alpha_1=0.5$  & 17.6 & 28.0 & 13.0 & 18.1 & 9.9  & 7.2  \\
\bottomrule
\end{tabular}
}
\caption{Forgetting (BWT) of MixLoRA-DSI on NQ320k under different combinations of $\alpha_1$ and $\alpha_2$.}
\label{tab:alpha_ablation_bwt}
\end{table}

\subsection{Ablation studies on the auxiliary router loss}
Table~\ref{tab:ablation_router_aux_loss} provides a breakdown of the contribution of each term in the auxiliary loss:

Regarding NQ320k, using both terms achieves significantly lower BWT (i.e., less forgetting) while still improving AP (Average Performance) compared to not using any. In contrast, using only one term improves AP at the cost of exacerbating BWT (i.e., more forgetting). For MSMARCO, using both terms also achieves significantly lower BWT with only a minor impact on AP. In contrast, using only one term either decreases AP or increases BWT. In conclusion, both loss terms in the proposed auxiliary loss contribute meaningfully to achieving a balance stability–plasticity trade-off.

\begin{table}[t]
\scalebox{0.85}{
\setlength{\tabcolsep}{1.2pt}
\centering
\begin{tabular}{cc|cccc}

\toprule
\multicolumn{2}{c|}{Methods} & \multicolumn{2}{c}{\(\text{AP}_4 \uparrow\)} & \multicolumn{2}{c}{\(\text{BWT}_4 \downarrow\)} \\
{First term} & {Second term} & R@10 & M@10 & R@10 & M@10 \\

\midrule 
\xmark & \xmark & 36.2 & 20.7 & 33.1 & 38.7 \\
\cmark & \xmark & 44.2 & 25.2 & 35.8 & 49.7 \\
\xmark & \cmark & 52.5 & 32.8 & 15.8 & 21.1 \\ 
\cmark & \cmark & \underline{68.1} & \underline{55.2} & 13.0 & \underline{18.1} \\

\bottomrule

\end{tabular}
}
\caption{Ablation results on the contribution of each term in the proposed auxiliary router loss (Eq.~\ref{eq:router_aux_loss}).}
\label{tab:ablation_router_aux_loss}
\end{table}

\subsection{Ablation studies on optimization objective's hyperparameters}
\label{appendix:alpha_ablation}

We conduct experiments to study the impact of different $\alpha_1$ (i.e., the auxiliary router loss) and $\alpha_2$ (i.e., KL divergence to regularize the updates to the RQ embeddings) combinations in Equation~(\ref{eq:final_loss}) on NQ320k.

In practical continual learning settings, preserving past knowledge (lower BWT) is often as critical as learning new knowledge (higher AP). The results show a trade-off between AP  (Table~\ref{tab:alpha_ablation_ap}) and BWT (Table~\ref{tab:alpha_ablation_bwt}): generally, a larger $\alpha_1$ slightly boosts AP performance but lowers BWT, while a larger $\alpha_2$ negatively impacts AP but reduces BWT. Our default choice of $\alpha_1=1.0$ and  $\alpha_2=0.1$ strikes a balance between the two metrics, achieving the best AP while maintaining relatively low BWT on both Recall@10 and MRR@10.

\subsection{Additional results on LongEval}
As requested by the reviewers during the rebuttal and due to limited computational resources and time constraints, we conducted experiments on the LongEval Challenge~\cite{longeval2025} using 3 training epochs for selected methods. Following the DSI++ benchmark setup, we selected four timesteps (October 2022 to January 2023) to simulate a continual learning scenario. We could not use the latest LongEval 2025 test set, as it had not been released during the rebuttal period, nor could we access the 2024 test set.

Details of the dataset used in our experiments are summarized in Table~\ref{tab:longeval-dataset}. Each new split contains only newly added unique documents. The annotated queries are divided into an 80-20 train-test split, with only the annotated training queries from 2022-10 used to construct the RQ-based docids.

We compare our proposed methods against CLEVER, the strongest continual learning baseline for generative retrieval, in Table~\ref{tab:longeval_result}. Our findings on LongEval are consistent with trends observed on NQ320k and MSMARCO. Despite being rehearsal-free, MixLoRA-DSI\(^\text{-PT}\) achieves competitive performance relative to CLEVER (n = 512) with EWC regularization. Furthermore, pre-training the experts substantially boosts MixLoRA-DSI’s performance, yielding the best results in this setup.

\begin{table*}[t]
\setlength{\tabcolsep}{2.5pt}
\renewcommand{\arraystretch}{1.0}
\small
\centering
\scalebox{0.9}{
\begin{tabular}{lcccccccccc}

\toprule

\multicolumn{8}{c}{\textbf{LongEval (Recall@10/MRR@10)}} \\

\textbf{Method} & \(\mathbf{D_0}\uparrow\) & \(\mathbf{D_1}\uparrow\) & \(\mathbf{D_2}\uparrow\) & \(\mathbf{D_3}\uparrow\) & \(\text{AP}_3\uparrow\) & \(\text{BWT}_3\downarrow\) & \(\text{FWT}_3\uparrow\) \\ 

\midrule

CLEVER(n=512) & 5.7/3.2 & 6.8/4.2 & 6.2/4.5 & 9.9/8.5 & 7.1/5.1 & 2.0/4.9 & 8.9/8.3 \\
CLEVER(n=1024) & 6.4/5.0 & 8.6/7.4 & 6.8/5.9 & 9.5/9.1 & 7.8/6.8 & 1.3/2.7 & 9.1/8.5 \\
MixLoRA-DSI\(^{\text{-PT}}\) & 6.4/6.7 & 8.4/8.5 & 7.0/7.5 & 9.3/9.4 & 7.3/6.0 & \textbf{1.1/1.8} & 8.2/6.4 \\
MixLoRA-DSI & \textbf{10.5/11.5} & \textbf{11.1/11.5} & \textbf{10.2/10.9} & \textbf{11.7/11.6} & \textbf{10.9/11.4} & 1.4/2.2 & \textbf{11.8/12.4} \\

\bottomrule

\end{tabular}
}
\caption{Models' performance after indexing LongEval's \(D_3\).}
\label{tab:longeval_result}
\end{table*}

\section{More details about residual quantization-based docids}
\label{appendix:rq_docids}

We adopt Residual Quantization (RQ)-based docid~\cite{zeng_ripor_www24}. RQ-based docid is the first docid that is effective for large-scale standard IR datasets~\cite{bajaj_msmarco_arxiv16}.
while most GR works have only shown competitive performance on smaller-scale benchmarks~\cite{pradeep_genirscaling_emnlp23}. 
For each document \(d \in D\), we obtain its document embedding by treating the pre-trained backbone of DSI as a dense encoder:
\begin{equation}
    \label{eq:doc_vector}
    e_d=\text{Decoder}(\text{<bos>}; \text{Encoder}(d)) \in \mathbb{R}^{\text{dim}}.
\end{equation}
After obtaining the set of document embeddings \(E_D=\{e_{d_1}, \ldots, e_{d_{|D|}}\}\), we train RQ with \(M\) codebooks, each containing \(K\) centroids (i.e., codewords), where the \(m\)-th codebook is defined as \(C^m \coloneqq \{c_k\}_{k=1}^{K} \in \mathbb{R}^{\text{dim}\times K}\). We further denote the \(k\)-th centroid in the \(m\)-th codebook as \(C^m[k] \in \mathbb{R}^{\text{dim}}\). The RQ codebooks are optimized to approximate the document embeddings:
\begin{equation}
    \label{eq:rq_docid}
    e_d \approx \sum_{m=1}^{M} C^{m}[i_m], \quad i_m \in [1, K].
\end{equation}
The trained RQ codebooks is used to generate an RQ code of length \(M\) for each document \(d\). The learned RQ centroids are concatenated to the DSI's vocabulary weights \(W_{vocab} \in \mathbb{R}^{\text{dim}\times |V|}\) with length \(|V|\)  for subsequent optimization: \(W_{\text{RQ}}=\{W_\text{vocab}; C^{1}; \ldots; C^{M}\} \in \mathbb{R}^{\text{dim}\times(|V|+M*K)}\), where \(\{\cdot;\cdot\}\) is concatenation.
\section{MixLoRA: Mixture of LoRA Experts}
\label{appendix:mixlora}
In this section, we elaborate the our implementation of MixLoRA in MixLoRA-DSI. 
We extend the MoE formulation in Section~\ref{subsec:moe} by introducing LoRAs as experts, namely MixLoRA. We replace selected Feed-Forward Network (FFN) layers of DSI with MixLoRA. 
Consider a T5-based~\cite{raffel_t5_jmlr20} DSI model, the Feed-Forward Network (FFN) is a two-layer Multi-Layer Perceptron (MLP) that processes an input \(x\). For simplicity, we omit dropout and activation functions without loss of generality:
\begin{equation}
    \label{eq:t5_ffn}
    \text{FFN}(x) = x + W_\text{out}(
        W_\text{in} (\text{LN}(x))
    ),
\end{equation}
where \(\text{LN}(\cdot)\) denotes layer normalization~\cite{ba_layernorm_arxiv16}; \(W_\text{in}\) and \(W_\text{out}\) are the input and output weight matrix, respectively. In MixLoRA, each layer of the FFN is adapted with a corresponding set of LoRAs\footnote{Compared to~\citet{yu_moeloraclvision_cvpr24}, our formulation uses two unique sets of LoRAs for each FFN layer}, while keeping the original layer weight frozen.
Let \(\mathbb{L}_{\text{in}}=\{\Delta_\text{in}^{i}\}_{i=1}^{N}\) and \(\mathbb{L}_{\text{out}}=\{\Delta_\text{out}^{i}\}_{i=1}^{N}\) denote the sets of \(N\) LoRAs with \(N \geq 2 \) for adapting \(W_\text{in}\) and \(W_\text{out}\), respectively.
The output of adapting \(W_\text{in}\) is:
\begin{equation}
    x^\prime = W_\text{in}(\text{LN}(x)) + \sum\limits_{i=1}^{N} Topk(p_i(x))\Delta_{\text{in}}^{i}(x),
\end{equation}
The final output of MixLoRA is computed as:
\begin{equation}
    \label{eq:mixlora_moe}
    \begin{gathered}
    \text{MixLoRA}(x) = \\
    x + W_\text{out}(x^\prime) + \sum\limits_{i=1}^{N} Topk(p_i(x)) \Delta_{\text{out}}^{i}(x^\prime).
    \end{gathered}
\end{equation}
By sharing the gate values across both MLP layers, we avoid the need for another router.
\section{Implementation details}
\label{appendix:implementation_details}

All experiments are conducted on a single NVIDIA A100 80GB GPU, using the AdamW optimizer~\cite{loshchilov_adamw_arxiv17} with a weight decay of 0.01 and learning rate of \(1e^{-3}\). We use linear learning rate schedule with 10\% warm-up steps, and gradient norm clipping of 1.0. We use T5-base~\cite{raffel_t5_jmlr20} as the backbone for all generative retrieval (GR) models, initializing pre-trained checkpoints from Huggingface~\cite{wolf_transformers_emnlp20}. The maximum input sequence length is 256. For constrained beam search inference, we use a beam size of 10 and decode for up to 8 steps.

The initial checkpoint is prepared following the first two steps of RIPOR’s implementation~\cite{zeng_ripor_www24} docid initialization and Seq2Seq pre-training. We construct RQ-based docids using RQ codebooks with \( M=8 \) and \( K=2048 \) centroids per codebook. The initial checkpoint \( M_0 \) is trained with a batch size of 1024 and a learning rate of 0.001. Training runs for 20K steps on NQ320k and 250K steps on MSMARCO.

\subsection{Traditional IR Models}
\begin{itemize}
    \item \textbf{BM25:} We use FAISS~\cite{johnson_faiss_tbd19} to index \( D_0 \) and retrieve documents using Pyserini’s BM25 implementation~\cite{lin_pyserini_sigir21}.
    \item \textbf{DPR:} We train a BERT-based dual-encoder DPR~\cite{devlin_bert_naacl19,karpukhin_dpr_emnlp20} on \( D_0 \) and use it for indexing and retrieval on new corpora with FAISS. The implementation follows~\citet{gao_gcdpr_acl21}, using the BERT-base checkpoint from Huggingface.
\end{itemize}

\subsection{Generative Retrieval Models}
We evaluate various GR models in our rehearsal-free dynamic corpus setting. 
During continual indexing, we train all models with with a learning rate of 0.001. We use a batch size of 128 and 512 for NQ320k and MSMARCO, respectively.
Starting from \( M_0 \), 
GR and Non PEFT-based CLGR models are trained for 5 epochs on NQ320k and MSMARCO;
PEFT-based CLGR methods and PEFT are trained for 10 epochs on NQ320k and 5 epochs on MSMARCO.

\begin{itemize}
    \item \textbf{BASE:} We reproduce using the code from~\citet{zeng_pag_sigir24,zeng_ripor_www24}.
    \item \textbf{DSI++:} We exclude generative replay, as it requires training an additional model on \( D_0 \), violating the rehearsal-free constraint. Since the source code is unavailable, we implement SAM~\cite{foret_sam_2020} using an open-source repository\footnote{\url{https://github.com/davda54/sam}}.
    \item \textbf{CLEVER:} We experiment with previous sample sizes \( n=512 \) and \( n=1024 \) for EWC regularization. We implement EWC based on open-source materials\footnote{\url{https://github.com/ContinualAI/colab/blob/master/notebooks/intro_to_continual_learning.ipynb}}.
    \item \textbf{CorpusBrain++:} We employ bottleneck adapters\footnote{\url{https://docs.adapterhub.ml/methods.html\#bottleneck-adapters}} to all encoder and decoder blocks of T5. The reduction factor is set to 96 to match the rank-8 LoRAs used in MixLoRA-DSI.
    \item \textbf{PromptDSI:} since the source code is not released, we implement it according to the paper. Neural topic embeddings are obtained from BERTopic~\cite{grootendorst_bertopic_arxiv22}, using \( M_0 \) as the encoder to prevent data leakage during continual indexing.
\end{itemize}

\subsection{MixLoRA-DSI}
For continual indexing, we use LoRAs with rank 8, a dropout rate of 0.05, and a scaling factor of 16. We build on an open-source MoE implementation\footnote{\url{https://github.com/lucidrains/st-moe-pytorch}}. We find \(\alpha_1 =1, \alpha_2=0.1\) to be robust (Appendix~\ref{appendix:alpha_ablation}). Regarding slow-learner in the RQ-based docids embeddings, we scale down the gradient of concatenated RQ token weights by a factor of 100.

Following insights from prior PEFT studies~\cite{wang_dualprompt_eccv22,huynh_promptdsi_arxiv24}, we apply MixLoRA to the first five layers of the decoder, as it is responsible for autoregressive decoding. By default, MixLoRA employs a top-2 router, initializing with two unique sets of LoRAs, discussed in Appendix~\ref{appendix:mixlora}. The token-wise energy score OOD thresholds \(\tau\) are set to the average exponential moving average of energy scores.

For NQ320k, all pre-trained MixLoRA-DSI variants use a batch size of 32, while other variants use a batch size of 128. All models are trained for 10 epochs. The layer-wise OOD query threshold \(\delta\) is set to 1\% for pre-trained variants and 5\% for MixLoRA-DSI\(^{\text{PT}}\).

For MSMARCO, MixLoRA-DSI variants use a batch size of 512 and are trained for 5 epochs. The layer-wise OOD query threshold \(\delta\) is set to 0.01\% for pre-trained variants and 10\% for MixLoRA-DSI\(^{\text{PT}}\).

The naive expansion baseline follows the same setup as MixLoRA-DSI but without pre-training on \(D_0\), expanding naively, and using the original MoE router with original load balancing loss.

\end{document}